\documentclass[fleqn,10pt]{wlscirep}
\usepackage[utf8]{inputenc}
\usepackage[T1]{fontenc}
\usepackage{bm}
\usepackage{upgreek}
\usepackage{amsmath}
\usepackage{amssymb}

\title{Fiber-taper collected emission from NV centers in high-$Q/V$ diamond microdisks}

\author[1]{Tamiko Masuda}
\author[1,2]{J.P.E. Hadden}
\author[1,3]{David P. Lake}
\author[1]{Matthew Mitchell}
\author[1]{Sigurd Fl\aa gan}
\author[1,*]{Paul E. Barclay}
\affil[1]{Department of Physics and Astronomy and Institute for Quantum Science and Technology, University of Calgary, Caglary, AB, T2N 1N4, Canada}
\affil[2]{Currently with the School of Engineering, Cardiff University, Cardiff, CF24 3AA, United Kingdom}
\affil[3]{Currently with the Applied Physics Department, California Institute of Technology, Pasadena, CA 91125, USA}

\affil[*]{e-mail: pbarclay@ucalgary.ca}

\begin{abstract}
Fiber-coupled microdisks are a promising platform for enhancing the spontaneous emission from color centers in diamond. The measured cavity-enhanced emission from the microdisk is governed by the effective volume ($V$) of each cavity mode, the cavity quality factor ($Q$), and the coupling between the microdisk and the fiber. Here we observe photoluminescence from an ensemble of nitrogen-vacancy centers into high $Q/V$ microdisk modes, which when combined with coherent spectroscopy of the microdisk modes, allows us to elucidate the relative contributions of these factors. The broad emission spectrum acts as an internal light source facilitating mode identification over several cavity free spectral ranges. Analysis of the fiber-taper collected microdisk emission reveals spectral filtering both by the cavity and the fiber-taper, the latter of which we find preferentially couples to higher-order microdisk modes. Coherent mode spectroscopy is used to measure $Q\sim 1\times10^{5}$ -- the highest reported values for diamond microcavities operating at visible wavelengths. With realistic optimization of the microdisk dimensions, we predict that Purcell factors of $\sim 50$ are within reach.
\end{abstract}
\begin{document}

\flushbottom
\maketitle

\thispagestyle{empty}

\section{Introduction}
Color centers in diamond are luminescent defects consisting of vacancies and substitutional impurity atoms embedded in the carbon lattice. These color centers often combine long spin coherence times\,\cite{Balasubramanian2009,Abobeih2018,Bradley2019,Stas2022} with spin-selective optical transitions\,\cite{Tamarat2008,Robledo2011NJP,Muller2014}.
The ability to perform all-optical spin control\,\cite{Rogers2014PRL,Chu2015,Becker2018} and single-shot spin readout\,\cite{Robledo2011,Sukachev2017,Irber2021} makes these color centers promising qubit candidates for photonic quantum architectures\,\cite{Johnson2017,Awschalom2018,Ruf2021a,Shandilya2022}. In particular, the optically addressable electron spin associated with the nitrogen-vacancy (NV) center\,\cite{Doherty2013} has been used in successful demonstrations of quantum memories\,\cite{Dutt2007,Fuchs2011,Bradley2019,Bradley2022,Abobeih2022}, spin-photon\,\cite{Togan2010} and spin-spin\,\cite{Bernien2013,Hensen2015,Kalb2017} entanglement, and quantum teleportation\,\cite{Pfaff2014}, culminating in the demonstration of multinode quantum networks\,\cite{Pompili2021,Hermans2022}. However, the ability to extend these proof-of-principle demonstrations to practical on- and off-chip quantum networks\,\cite{Kimble2008} requires the efficient generation and distribution of entanglement across remote network nodes\,\cite{Wehner2018}. The success rate of heralded entanglement schemes utilizing one-photon\,\cite{Hermans2022,Humphreys2018} and two-photon protocols\,\cite{Bernien2013,Pfaff2014,Hensen2015} are inherently limited by the collection probability of coherent photons. For experiments using NV centers, the flux of coherent photons is limited by the small branching into the zero-phonon line of $\sim3\,\%$ and a poor collection efficiency due to total internal reflection at the diamond-air interface\,\cite{Riedel2017}. In turn,  these shortcomings limit  scalability to systems consisting of many entangled nodes, such as quantum-repeater-based long-distance quantum communication\,\cite{Guha2015} and the realization of quantum networks\,\cite{Kimble2008,Wehner2018}.

The tight confinement of light in optical resonators combining a high quality factor, $Q$, with a small mode volume, $V$, leads to enhanced light-matter interactions\,\cite{Vahala2003,Wang2019,Najer2019}. For example, by utilizing the Purcell effect\,\cite{Purcell1946}, optical cavities have been used to enhance the photon flux from color centers in diamond\,\cite{Faraon2011,Faraon2012,Riedel2017,Ruf2021}. While coupling between diamond color centers and optical cavities have been demonstrated using a variety of different geometries, including hybrid platforms\,\cite{Fu2008,Barclay2009,Barclay2011,Gould2016,Schmidgall2018,Fehler2021,Riedel2023arXiv} and tunable Fabry-Perot microcavities\,\cite{Albrecht2013,Johnson2015,Kaupp2016,Riedel2017,Benedikter2017,Haussler2019,Jensen2020,Ruf2021,Bayer2022ArXiv}, realizing high $Q/V$ monolithic cavities directly from single-crystal diamond (SCD) allows for minimizing $V$ while maximizing the emitter-cavity coupling by maximizing the overlap of the emitter dipole-moment with the cavity mode. Furthermore, the use of monolithic resonators facilitate the implementation of on-chip photonic buses for efficient photon extraction\,\cite{Wan2020}. This, in conjunction with the fact that the highest quality color centers are found in bulk SCD\,\cite{Balasubramanian2009,vanDam2019,Kasperczyk2020,Chakravarthi2021,Yurgens2021,McCullian2022}, has spurred the invention of fabrication techniques to further develop SCD devices\,\cite{Burek2014,Burek2016,Appel2016,Sipahigil2016,Castelletto2017,Burek2017,Challier2018,Mitchell2019,Nguyen2019,Hedrich2020,Duan2021}.

At the visible wavelengths resonant with diamond color center optical transitions ($637\,\textrm{nm}$ for the NV center), SCD ring resonators produced by a thinned membrane technique have been reported with $Q\sim 3\times10^{4}$ and effective mode volume, $V_{\textrm{eff}}\sim 10^{1}\,(\lambda/n)^{3}$\,\cite{Faraon2011}. In this wavelength range, racetrack resonators\,\cite{Burek2014} fabricated using angled etching of SCD\,\cite{Burek2012} have demonstrated $Q\sim 6\times10^{4}$ -- the previously largest $Q$-factor reported for monolithic SCD resonators -- while open Fabry-Perot microcavities have been reported with $Q>10^5$\,\cite{Janitz2015,Flagan2022,Flagan2022Dres}. However, the increased length of these resonators limits the smallest achievable $V$. Photonic crystal nanocavities, on the other hand, have been demonstrated with $Q\sim 1\times 10^{4}$\,\cite{Burek2017}, and ultra-low $V_{\textrm{eff}}< (\lambda/n)^{3}$\,\cite{Nguyen2019,Bhaskar2020}. Recently, quasi-isotropic undercut etching\,\cite{Khanaliloo2015} has been used to create high $Q/V$ microdisks\,\cite{Khanaliloo2015NanoLett,Mitchell2016,Mitchell2019} and photonic crystal cavities\,\cite{Riedrich-Moller2012,Mouradian2017a}. While the latter have been realized at visible wavelengths, and coupling to color centers have been demonstrated\,\cite{Hausmann2013,Jung2019}, previous studies of diamond microdisks have been predominantly limited to telecommunication wavelengths and applications related to cavity optomechanics\,\cite{Mitchell2016,Lake2018,Mitchell2019a,Lake2020NatCom,Shandilya2021}.   

In this work, we demonstrate coupling between an ensemble of NV centers and whispering gallery modes (WGM) in a SCD microdisk. We further employ the NV center emission, combined with coherent mode spectroscopy, to simultaneously probe the coupling of the NV centers to the microdisk and the coupling between the microdisk and the fiber-taper. To this end, we use the fiber-taper waveguide to evanescently probe and collect the microdisk mode PL, and harness the broad phonon sideband (PSB, $600-800\,\textrm{nm}$) as an internal lightsource to perform broadband cavity mode spectroscopy over several cavity free spectral ranges (FSR). We present a quantitative comparison of these results to coherent fiber-taper transmission spectroscopy performed using tunable lasers with a smaller tuning range ($5\,\textrm{nm}$) but higher spectral resolution.

The subsequent analysis reveals that high-intensity PL signals are not necessarily indicative of modes that promise the largest Purcell enhancement. The transmission measurements demonstrate microdisk $Q$-factors as high as $\sim 1\times10^{5}$ with $V_{\textrm{eff}}$ expected to be $\sim10\,(\lambda/n)^{3}$ estimated from finite-difference time-domain (FDTD) simulations. With optimized dimensions, simulations predict that $V_{\textrm{eff}}\sim(\lambda/n)^{3}$ is within reach, suggesting that SCD microdisks may be able to match the Purcell factors demonstrated using photonic crystal\,\cite{Riedrich-Moller2012,Hausmann2013} and Fabry-Perot cavities\,\cite{Riedel2017,Jensen2020,Ruf2021}.

The remainder of this paper is organized as follows. We begin by detailing the fiber-taper--microdisk system, followed by an outline of the experimental procedure. Next, we study the fiber-taper-collected PL spectra by comparing the measured intensities to those expected from a simple model of the system, based on the $Q$-factors extracted from the fiber-taper transmission measurements. Following this, we use broadband FSR data to perform mode-family identification to gain further insight into the fiber-taper--microdisk system. Finally, in the outlook, we discuss optimization of the microdisk geometry with regards to Purcell enhancement of spontaneous emission from color centers embedded in the microdisks.

\section{Experimental setup: fiber and free-space coupling}

\begin{figure}[ht!]
	\centering\includegraphics[width=\textwidth]{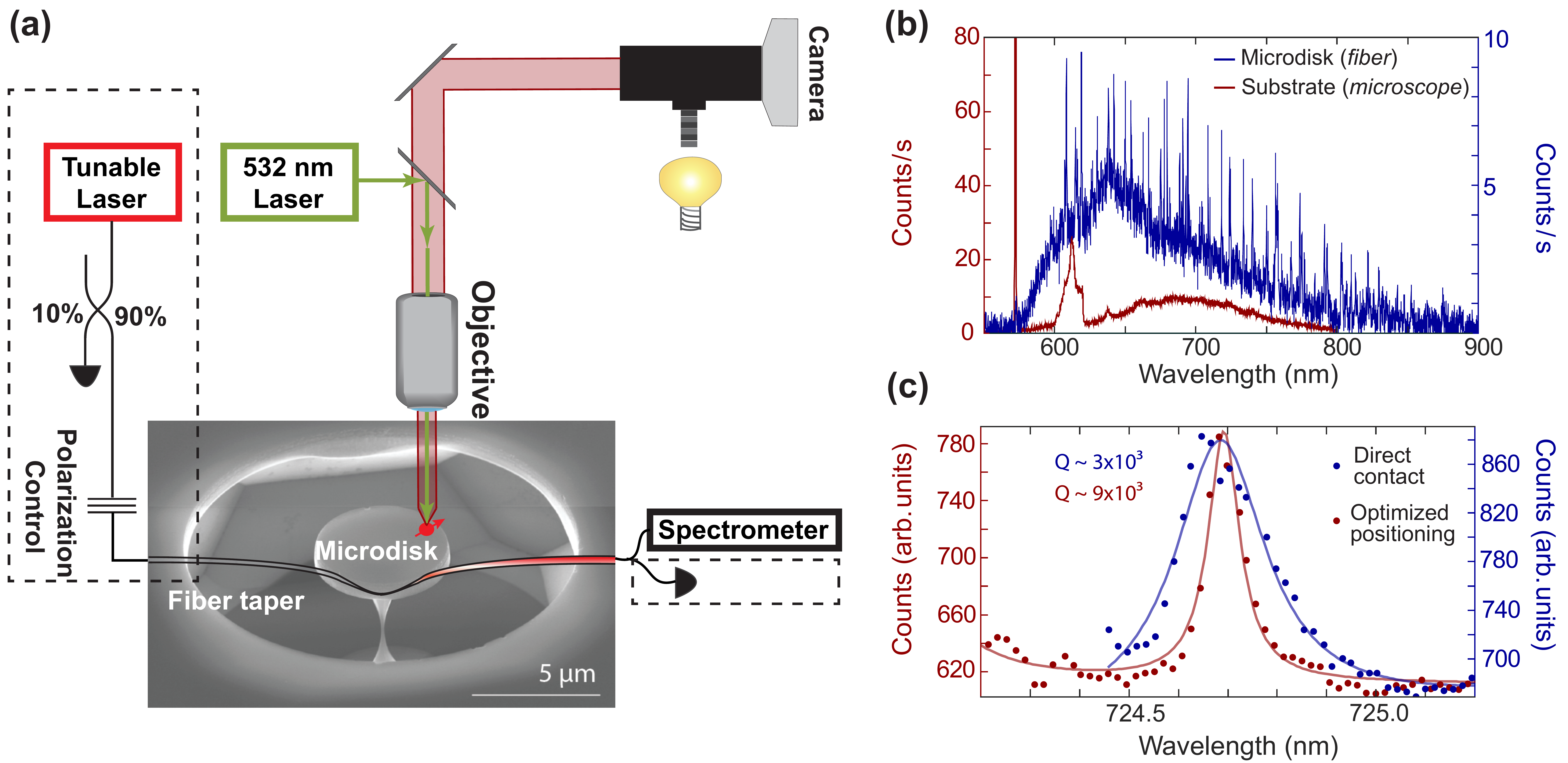}
	\caption{\textbf{(a)} A simplified schematic of the experimental setup used for fiber-taper collected PL and cavity transmission measurements. Included is a SEM image of one of the diamond microdisks studied in this work. The objective lens used for free-space excitation can be manually moved using an $xyz$ translation stage. \textbf{(b)} In blue: fiber-collected PL from a diamond microdisk of [$r,t$]$\sim$[$2.5\,\upmu\textrm{m}$,\,$850\,\textrm{nm}$]. In burgundy: reference spectrum showing the NV center PSB obtained via confocal microscope collection of NV center PL signal from an implanted electronic grade diamond substrate, taken with an alternative setup. \textbf{(c)} Fiber-taper collected PL spectroscopy for two fiber positions. Positioning the fiber carefully next to the device yield $Q\sim 9\times10^{3}$ (burgundy). When the fiber is touching the device (blue line), a threefold reduction of the $Q$-factor is observed, resulting in $Q\sim 3\times10^{3}$.
	}
	\label{fig:setup} 
\end{figure}

In the experiment presented in this work, we study SCD microdisks fabricated following the quasi-isotropic undercut etching process reported in Refs.\,\cite{Mitchell2016, Khanaliloo2015NanoLett,Mitchell2019}. The starting material were diamond chips with a $\langle 100 \rangle$ crystal orientation grown using  chemical vapor deposition (Element Six). These chips are ``optical grade'', with an estimated nitrogen impurity concentration of $<1\,\textrm{ppm}$. In brief, the microdisk fabrication procedure involves three main steps: (1) writing device patterns using electron beam lithography (EBL); (2) transferring patterns to a silicon nitride hard mask and subsequently to the diamond via anisotropic reactive ion etching; and (3) undercutting of the devices with a quasi-isotropic etch\,\cite{Khanaliloo2015,Khanaliloo2015NanoLett}. As shown in Fig.\ref{fig:setup}\,(a), the resulting microdisks are supported by thin pedestals and have radius $r\sim2-4\,\upmu\textrm{m}$ and thickness $t \sim0.8-1.0\,\upmu\textrm{m}$.

The experimental setup used to study both the fiber transmission and PL from the microdisks is shown schematically in Fig.\,\ref{fig:setup}\,(a). The fiber-taper is positioned within the near-field of the microdisks using stepper-motor translation stages (Suruga Seiki), while imaged with white light on a CCD camera. When coupling the fiber-taper to the microdisk, it is important to avoid interactions with the unpatterned substrate surrounding the ring-shaped windows defining the microdisks. Therefore, the fiber-taper is produced with a \textit{dimple}\,\cite{Michael2007}. The fabrication procedure of the dimpled fiber-taper goes as follows. First, a single-mode optical fiber (Nufern 630-HP) designed for operation at visible wavelengths is heated using a stationary (no dither) hydrogen torch setup and stretched to produce a fiber-taper that is single mode at $\sim640\,\textrm{nm}$. Ideally, this corresponds to a fiber-taper radius of $a\sim230\,\textrm{nm}$\,\cite{Yariv2006}. The dimple is then created by annealing the taper while it is wrapped around a ceramic knife edge. The dimpled fiber-taper is mounted in a U-bend shape, which provides mechanical stability and resilience.

To measure the transmission through the fiber-taper, we sweep the output of a narrow linewidth tunable diode laser (Newport Velocity, $\lambda=735-740\,\textrm{nm}$, $\delta\nu\leq200\,\textrm{kHz}$), while monitoring the transmitted signal on a photodiode. To maximize the contrast of the observed cavity resonances, the polarization of the incoupled light is adjusted using a fiber paddle controller. Wavelength- and time-dependent fluctuations in the laser output can make it difficult to discern low-contrast features in the transmission measurement. To mitigate this effect, a tap-off of the laser output provides a means to remove the laser-related features from the transmitted signal, thus aiding the identification of microdisk resonances.

For the fiber-taper-collected PL measurements, an ensemble of NV centers intrinsic to the optical grade diamond are excited from the top using a $532\,\textrm{nm}$ laser with power on the order of mW using a microscope objective of numerical aperture ($\textrm{NA}=0.55$). The resulting PL from the NV center PSB provides an internal cavity light-source that can be used to characterize the optical modes of the microdisk\,\cite{Riedel2020}. The microdisk-filtered PL is collected by the fiber-taper and detected by a spectrometer. An example spectrum collected using this system is shown by the blue curve in Fig.\,\ref{fig:setup}\,(b). Sharp features associated with coupling to individual microdisk modes are evident over a broad range extending from $600-800\,\textrm{nm}$. This broad signal extends over the same wavelength range as the reference spectrum (burgundy) of the NV center PSB\,\cite{Kaupp2013,Dolan2018}. The reference spectrum was measured through direct free-space confocal microscope collection of PL from implanted NV centers in an electronic grade sample, using an alternative setup. It represents the standard shape of a PL spectrum from a diamond sample excited by $532\,\textrm{nm}$ light, including NV emission and Raman signatures at $572\,\textrm{nm}$ and $\sim612\,\textrm{nm}$\,\cite{Riedel2020}. 
The shape of the fiber-taper-collected signal is dependent on factors including the position of the excitation laser and the fiber-taper, respectively. We therefore note that the reference spectrum should not be used to compare collection efficiencies, as any variation between the two spectra arises due to the different collection methods.

At the close proximity required for evanescent coupling, the fiber-taper is attracted to the microdisk, often resulting in contact between the fiber-taper and the disk\,\cite{Burek2014}. However, it is possible to mitigate completely this contact by taking advantage of the geometry of the window in the substrate that defines the microdisk (see Fig.\,\ref{fig:setup}\,(a)). While maintaining sufficient separation between the sample surface and fiber-taper to prevent attraction, the taper can be roughly positioned at the desired lateral location. Then, by lowering the fiber quickly,  contact between  the window edges and the fiber-taper prevents the fiber from touching the microdisk. The importance of careful positioning of the fiber-taper is illustrated in Fig.\,\ref{fig:setup}\,(c), where PL measurements were recorded for two different fiber-taper positions. An increase in the cavity linewidth, and corresponding reduction in $Q$-factor, is observed when the fiber is in contact with the microdisk (blue line in Fig.\,\ref{fig:setup}\,(c) ). Intuitively, this reduction in $Q$-factor can be explained by, upon contact, the fiber effectively becomes part of the cavity and induces parasitic loss\,\cite{Spillane2003}.

\section{Fiber-coupled NV center photoluminescence and coherent mode spectroscopy}
In this section, we use fiber-taper-collected PL from the NV centers as a means to characterize the microdisk optical mode spectrum over the broad PSB wavelength range. In an ideal case, NV center PL collected by the fiber-taper and imaged on the spectrometer would reveal the bare microdisk cavity mode spectrum, where the relative peak intensities are governed by the $Q\,/V$ ratio of the individual modes. However, in practice, this is not the case as fiber-taper-collected PL is affected by the coupling between the fiber-taper and the microdisk -- an effect that we elucidate in the analysis presented below.

Figure\,\ref{fig:PkHt} shows fiber-taper-collected PL and fiber-taper transmission measurements for two microdisks, labeled $A$ and $B$ with dimensions [$r,t$]$\sim$[$2.0\,\upmu\textrm{m},\,800\,\textrm{nm}$] and [$r,t$]$\sim$[$3.10\,\upmu\textrm{m},\,850\,\textrm{nm}$], respectively. The correlation between peaks in PL and dips in the transmission spectrum, confirms that the observed PL peaks are signatures of microdisk optical modes. However, the strongest fiber-collected PL signals do not, in general, correspond to the highest-$Q$ microdisk modes measured in transmission. Rather, the peaks in PL can be aligned with lower-$Q$ microdisk modes coupled more strongly to the fiber-taper. Below we analyze both transmission and PL data to determine the relative impact of the optical mode $Q$-factor and coupling parameters on the observed signals.

The fiber-taper-collected PL intensity, $I_{j}$, of microdisk mode $j$, can be predicted by modeling the fiber-taper--microdisk collection system using an input-output formalism. Mode $j$ is defined to have field amplitude $a_j$, total loss rate $\kappa_{j}$, and resonant frequency $\omega_j$\,\cite{Spillane2003}. The mode is excited at frequency $\omega$ by a source term $s_j$, whose strength depends on the emission of the NV PSB into mode $j$. Given external coupling rate $\kappa^e_{j}$ from the microdisk mode into the forward propagating (i.e.\ measured) mode of the fiber-taper, the collected PL intensity from mode $j$ is $I_{j}=|\sqrt{\kappa^{e}_{j}}a_{j}|^{2}$. The cavity mode equation of motion is\,\cite{Barclay2009OpticsExpress}:  
\begin{equation}
\begin{split}
\frac{da_{j}}{dt} = \left(i\Delta\omega_{j} - \frac{\kappa_{j}}{2}\right)a_{j} + s_j\,,
\end{split}
\label{eqn:dadt}
\end{equation}
where $\Delta\omega_j = \omega - \omega_j$. This can be solved for a steady-state value of $a_j$ to obtain:
\begin{equation}
I_{j}=\frac{\kappa^{e}_{j}|s_{j}|^2}{(\Delta\omega_{j})^2 + (\kappa_{j}/2)^2}\,.
\label{eqn:I}
\end{equation} 
%
Note that similar expressions can be derived for the symmetric and anti-symmetric modes of doublet resonances observed for the highest $Q$ modes of our microdisks\,\cite{Srinivasan2006} .

Equation\,\eqref{eqn:I} can be expressed in terms of $Q_j=\omega_j/\kappa_j$ and $Q_j^e=\omega_j/\kappa^e_j$ and evaluated on resonance to obtain a peak height of:
\begin{equation}
I_{j}(\Delta\omega_j=0)=4Q_j\frac{Q_j}{Q_j^e}\frac{|s_{j}|^2}{\omega_j}\,.
\label{eqn:Ifj}
\end{equation}
From this expression, the emission into the fiber-taper for mode $j$ can be enhanced by increasing $Q_j$ or by increasing the coupling strength to the fiber-taper, as determined by $Q_j/Q_j^e$. The ratio of $Q_j$ to $Q_j^e$ is responsible for the dramatic differences in peak heights observed in Fig.\,\ref{fig:PkHt}, which we will quantify further below.

\begin{figure}[ht!]
	\centering	
	\includegraphics[width=\textwidth]{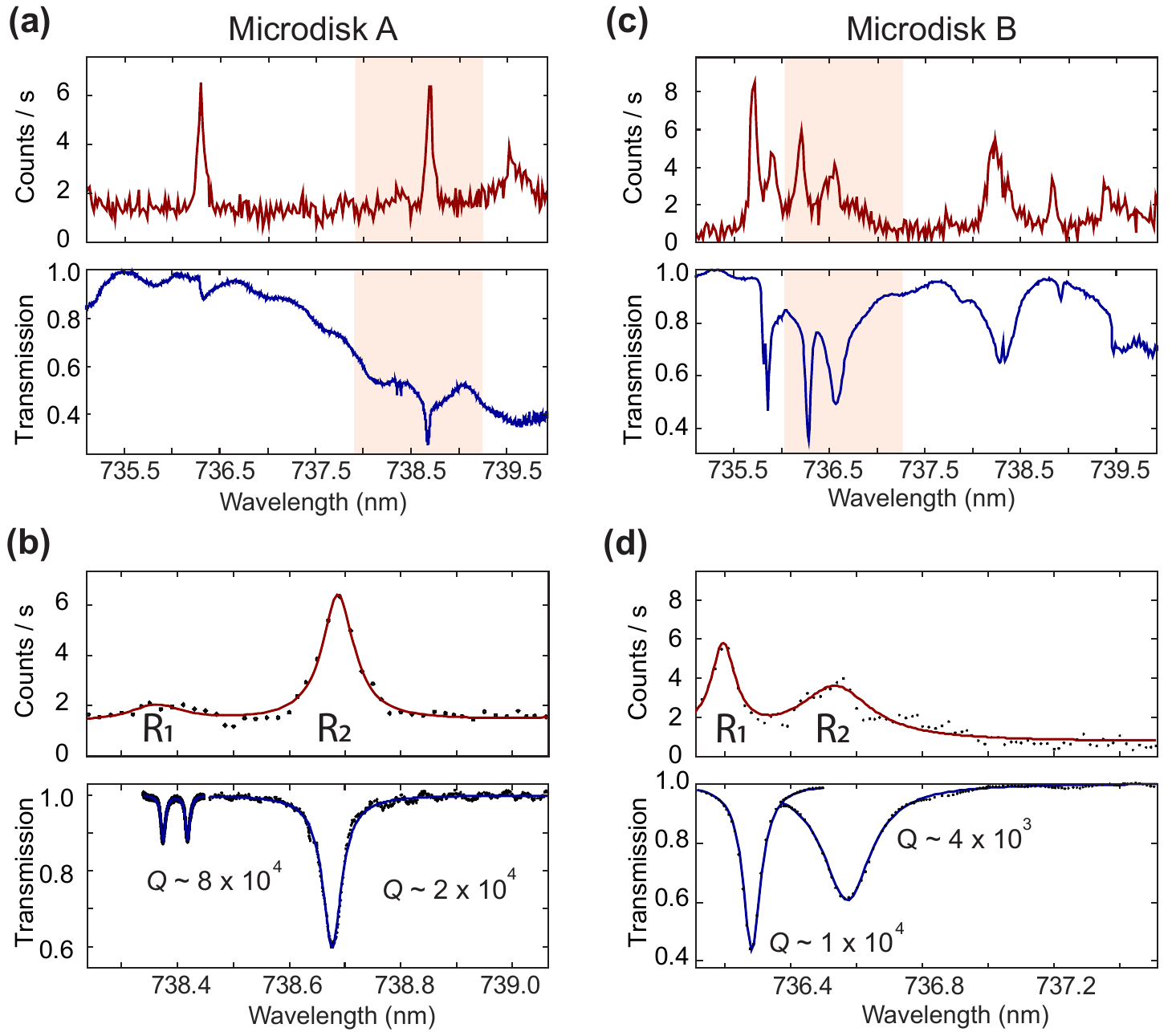}
	\caption{Transmission (blue) and photoluminescence (burgundy) measurements for microdisk $A$\,(a,\,b) with [$r,t$]$\sim$[$2.0\,\upmu\textrm{m}$,\,$800\,\textrm{nm}$] and microdisk $B$\,(c,\,d) with [$r,t$]$\sim$[$3.1\upmu\textrm{m}$,\,$850\,\textrm{nm}$]. Panel (a) and (c) show the measurement over a broad WL range. The  shaded regions are expanded to produce the fits displayed in panel (b) and (d). For the transmission measurements, the fit for each resonance have been performed independently, taking into consideration the modulated background, and superimposed in the figure. Discrepancies in PL and transmission data regarding resonance center wavelength are attributed to  slight calibration errors ($<0.1\,\textrm{nm}$).}
	\label{fig:PkHt}
\end{figure}

Microdisk $A$ supports a high $Q\sim 80\,000$ mode alongside lower $Q$ modes, as shown in the laser transmission measurements in Fig.\,\ref{fig:PkHt}\,(a) and Fig.\,\ref{fig:PkHt}\,(b).  Microdisk $B$, on the other hand, has a more complicated mode spectrum, as shown in Fig.\,\ref{fig:PkHt}\,(c) and \ref{fig:PkHt}\,(d). In Fig.\,\ref{fig:PkHt}\,(b), we see that mode $R_1$ of microdisk $A$ has $Q$-factor approximately 4 times larger than that of mode $R_2$. However, the PL associated with $R_1$ is barely discernible, while $R_2$ exhibits a strong peak. On the other hand, Fig.\,\ref{fig:PkHt}\,(d) shows that for microdisk $B$, the higher-$Q$ mode $R_1$ does have a larger peak height  in PL compared to the lower-$Q$ $R_2$ mode. The data illustrates that the peak height is not always an accurate measure of the $Q$-factor, and that variations in the peak height can be strongly influenced by differences in $Q_j^e$ and $s_{j}$. 

To quantify further the impact of the different contributions to the peak height, we define the \textit{peak height ratio}, $\alpha$, for a given pair of modes: 
\begin{equation} \label{eqn:alpha1}
\alpha = \frac{I_{\textrm{R}_1}(\omega_{\textrm{R}_1})}{I_{\textrm{R}_2}(\omega_{\textrm{R}_2})}\,.
\end{equation}
Table\,\ref{tbl:alpha} compares three distinct versions of $\alpha$, calculated under different assumptions to reveal the effect of $Q_j$ and $Q_j^e$ on the relative peak height. Here, $\alpha_{\textrm{I}}$ is the ratio of the measured PL intensity values. We next use Eq.\,\ref{eqn:Ifj} to calculate $\alpha$ for two different scenarios using the values of $Q_j$ extracted from the laser transmission measurement in  Fig.\,\ref{fig:PkHt}\,(b) and Fig.\,\ref{fig:PkHt}\,(d). First, assuming equal fiber-taper coupling for both modes, i.e.\ $Q_{\textrm{R}_1}^e=Q_{\textrm{R}_2}^e$, we calculate the Purcell enhanced peak intensity and the corresponding $\alpha_{\textrm{P}}$. Second, to account for the different fiber-taper coupling strength for mode $j$, we calculate the peak intensity, and the corresponding $\alpha_{\textrm{F}}$, by extracting $Q_j^e$ from the transmission resonance contrast and the measured $Q_j$. We note that in calculating $\alpha_{\textrm{P}}$ and $\alpha_{\textrm{F}}$, we made the assumption that $s_{\textrm{R}_1}=s_{\textrm{R}_2}$; a reasonable assumption on the grounds that the measurements are performed at room temperature where the microdisk--NV center system is in the ``bad emitter regime'', so long as modes $R_1$ and $R_2$ have comparable mode volume, which we revisit shortly. In this regime, emitters behave as a white light source\,\cite{Auffeves2009,Meldrum2010,Kaupp2016}, and thus $|s_j|^2$ is expected to be proportional to the intensity of the PSB at $\omega_j$, approximated to be  constant over the narrow wavelength ranges studied in Fig.\,\ref{fig:PkHt}\,(b) and  Fig.\,\ref{fig:PkHt}\,(d).

\begin{table}[htb!]
	\centering
	\caption{Peak height intensity ratios and related $Q$-factors for the modes defined in Fig.\,\ref{fig:PkHt}.}
	\begin{tabular}{c|c|c|c|c|c|c|c}
	\hline
		Microdisk & $Q_{\textrm{R}_1}$ & $Q_{\textrm{R}_2}$ & $Q^e_{\textrm{R}_1}$ & $Q^e_{\textrm{R}_2}$ & $\alpha_{\textrm{P}}$ & $\alpha_\textrm{F}$ & $\alpha_{\textrm{I}}$ \\
		\hline\hline
		 $A$ & $8\times10^{4}$ & $2\times10^{4}$	& $2.5\times10^6$ & $1.6\times10^5$ & $21$	& $1.3$ & $0.1$\\ 
		\hline
		 $B$ & $1\times10^{4}$ & $4\times10^{3}$	& $6.8\times10^4$ & $3.7\times10^4$ &	$7.6$		& $4.2$ & $1.8$\\ 
		\hline
	\end{tabular}
	\label{tbl:alpha}
\end{table}

For microdisk $A$,  the cavity transmission spectrum, used to calculate $\alpha_{\textrm{P}}$, predicts $I_{\textrm{R}_1}$ to be $21$ times larger compared to $I_{\textrm{R}_2}$. However, after considering the fiber-taper coupling strength, we predict that $I_{\textrm{R}_1}$ should be of similar magnitude ($1.3$) as $I_{\textrm{R}_2}$, as quantified by $\alpha_\textrm{F}$. This demonstrates how the microdisk-fiber-taper coupling strength, quantified by the ratio $Q_j/Q^e_j$, can compensate for a lower $Q_{j}$ when determining PL peak height\,\cite{Spillane2003}. Furthermore, the order of magnitude reduction from $\alpha_{\textrm{F}}$ to $\alpha_{\textrm{I}}$ points towards other factors affecting the system. For comparison, a similar, but less drastic tendency, is observed for Microdisk $B$.

We next elucidate on possible explanations accounting for the observed variations in $\alpha_{\textrm{I}}$. For microdisk $A$, the discrepancy between $\alpha_{\textrm{F}}$ and $\alpha_{\textrm{I}}$ can be attributed to limited spectrometer resolution -- the peak in PL is not fully captured by the spectrometer owing to the high $Q$-factor and doublet nature of mode $R_1$. Note that by sending the excitation laser straight to the spectrometer, we estimate the spectrometer resolution to be $\delta\lambda\simeq15\,\textrm{pm}$, corresponding to a spectrometer limited $Q_{\textrm{S}}=5\times10^4$ -- cavity resonances with $Q$-factors exceeding $Q_{\textrm{S}}$ can not be reliably resolved\,\cite{Barclay2009}. Furthermore, a non-unity $|s_{\textrm{R}_1}/\,s_{\textrm{R}_2}|$ would be manifested by variations in  $\alpha_{\textrm{I}}$. However, the present data is insufficient to conclude whether or not our assumptions for $s_j$ are valid. In particular, rather than approximating the PSB to be a white light source, one could model the NV center emission rate into each mode, as demonstrated for a single NV center coupled to a Fabry-Perot cavity\,\cite{Albrecht2013}. A closer agreement between $\alpha_{\textrm{F}}$ and $\alpha_{\textrm{I}}$ is found for Microdisk $B$. However, a disagreement is still present -- the factor of $\sim2$ difference between $\alpha_{\textrm{F}}$ and $\alpha_{\textrm{I}}$ can be explained by differences in the mode volume, as will be discussed further below.

\section{Microdisk mode identification and impact on the fiber-coupled NV center emission}
To better understand the coupling between the fiber-taper and the microdisk, we next identify microdisk WGM families using individual PL resonances. We start by identifying the cavity mode families based on the free spectral range. Next, we elucidate how modes of different order couple to the guided mode in the fiber-taper.

\subsection{The cavity free spectral range} 
Microdisk mode families are classified by TE\,(TM)$_{p,q}$, for predominantly transverse electric (magnetic) polarization and WGM radial (vertical) order number $p$ ($q$). For a given polarization and $(p,q)$, there exists a discrete spectrum of modes with a range of allowed azimuthal order number $m$. The FSR, defined as the spacing between mode $m$ and mode $m+1$ of the same family, can be estimated according to
\begin{equation}
\textrm{FSR}(m)=2\pi R n_\text{eff} \bigg(\frac{1}{m} - \frac{1}{m+1}\bigg)\,,
\label{eqn:fsr}
\end{equation}
where $n_\text{eff}$ is the effective refractive index, dependent on the mode family and varying slowly with $m$. Modes of different families can be distinguished by measuring $\textrm{FSR}(\lambda_j)$ of regularly spaced resonances of wavelengths $\lambda_j$ and determining $n_\text{eff}$. Comparison with simulated values of FSR can then be used to determine the polarization and ($p, q$) of each mode family.

\begin{figure}[t!]
	\centering\includegraphics[width=\textwidth]{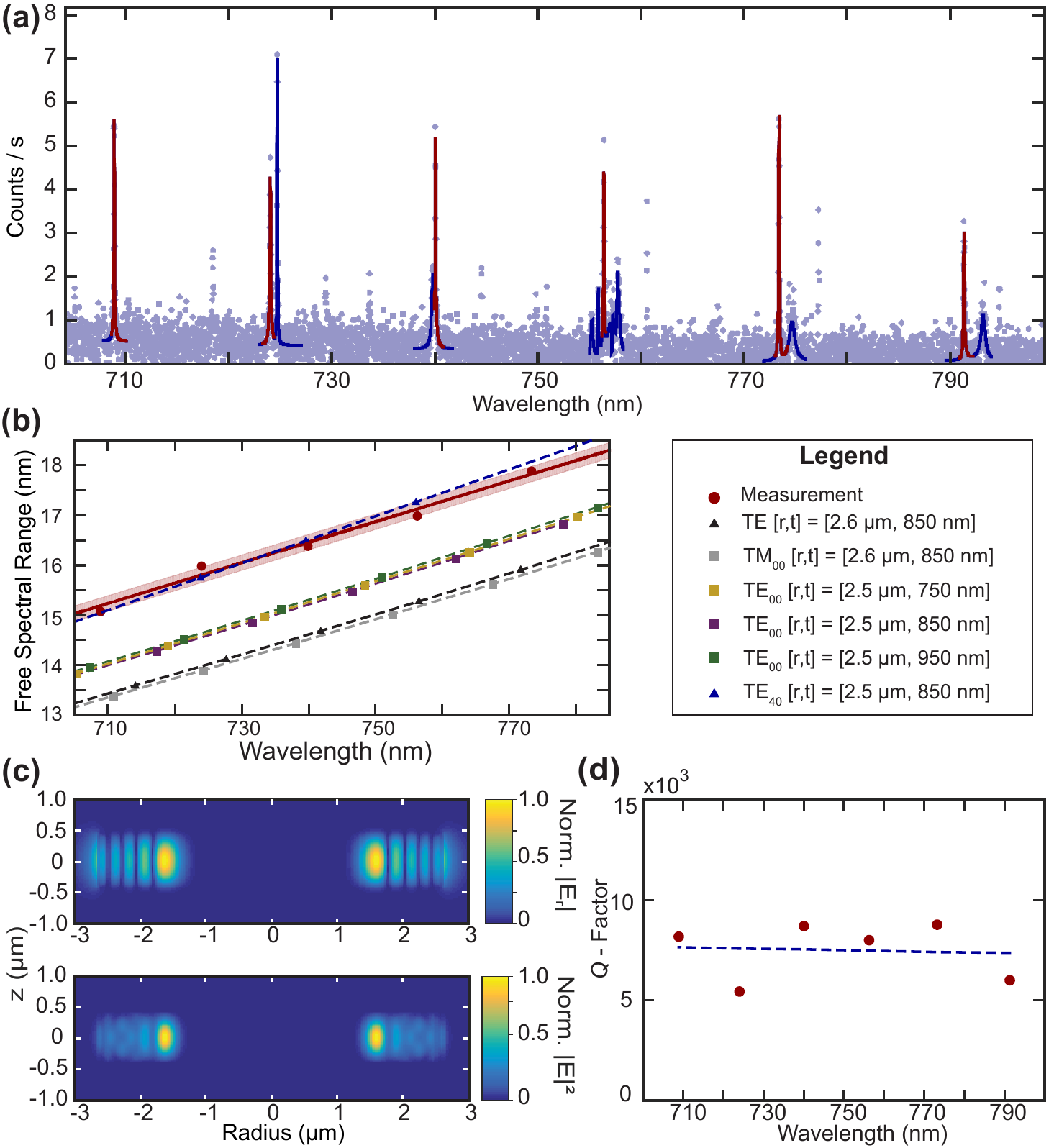}
	\caption{\textbf{(a)} PL from microdisk $C$ with dimensions [$r,t$]=[$2.6\,\upmu\textrm{m}$,\, $850\,\textrm{nm}$]. The blue lines show multiple Lorentzians fitted to the cavity modes. The burgundy lines are hand-selected resonances belonging to the same mode family, based upon the FSR. \textbf{(b)} Comparison between measured and simulated $\textrm{FSR}(\lambda_{m})$. The burgundy line shows a linear fit to the measured FSR from (a). The shaded area accounts for the uncertainty in the fit, calculated from the standard deviation of the residuals. \textbf{(c)} Simulated electric field profile for the TE$_{40}$ mode. Top panel: electric field in the radial direction, $E_{\textrm{r}}$. Bottom panel: total electric field squared, $|E|^2$. \textbf{(d)} $Q$-factor for the burgundy resonances in (a) as a function of wavelength.}
	\label{fig:fsr}
\end{figure}

In general, mode family identification for the relatively thick microdisks studied here can be challenging, in part due to their multi-mode nature and subsequently dense mode spectra. Such identification was not practical from the data collected for microdisks $A$ and $B$. 
However, a sparse spectrum shown in Fig.\,\ref{fig:fsr}\,(a)  owing to an advantageously positioned fiber-taper when measuring a third device with similar dimensions, microdisk $C$ ([$r,t$]=[$2.6\,\upmu\textrm{m}$,\,$850\,\textrm{nm}$]), lends itself to this analysis.

Figure\,\ref{fig:fsr}\,(a) shows the cavity mode spectrum for microdisk $C$. The measured FSR for the dominant resonances in the spectrum are plotted in Fig.\,\ref{fig:fsr}\,(b), along with the FDTD simulated FSR values for several different mode families. These simulations were obtained using the MEEP open-source software package\,\cite{Oskooi2010}. The measured FSR values are significantly larger compared to the simulated FSR values of the fundamental TE$_{00}$ and TM$_{00}$ modes. Accounting for an uncertainty in the radius on the order of $\sim10\,\textrm{nm}$ (EBL resolution) and an uncertainty in the thickness $\sim100\,\textrm{nm}$ (defined during the etching process and estimated via SEM), does not explain this difference. This is illustrated by the simulated values in Fig.\,\ref{fig:fsr}\,(b) accounting for a range of geometries. The high-intensity resonances in Fig.\,\ref{fig:fsr}\,(a), therefore, do not appear to be fundamental modes.

As alluded to by the typically dense mode spectra, simulations confirm that, for the dimensions of these microdisks, high-$Q$ modes of both higher radial and vertical order exist. As the radial order increases, $n_{\textrm{eff}}$ decreases and FSR as a function of wavelength increases. Consistent with these trends, the FSR of the TE$_{40}$ mode is found to be of similar magnitude to the measured values (Fig.\,\ref{fig:fsr}\,(b)). Note that, the TE$_{40}$ family has simulated radiation limited $Q> 10^{5}$ over the entire measured range -- a trend consistent with the measured $Q$-factors in Fig.\,\ref{fig:fsr}\,(d), where the average $Q_{\textrm{avg}}=7.4\times10^{3}$ shows no clear wavelength dependence, thus indicating that the measured modes are not radiation limited\,\cite{Borselli2004,Borselli2005}. Furthermore, the larger mode volume associated with higher-order TE$_{40}$ modes is consistent with the difference in the observed $\alpha_{\textrm{I}}$ ratios -- as shown below, the TE$_{40}$ mode have $V$ $\sim3$ times larger compared to the fundamental TE\,(TM)$_{00}$ modes.
\begin{table}
	\centering
	\caption{Comparison of fundamental TE\,(TM)$_{00}$ and TE$_{40}$ modes of radial order $m$ for select resonances near $740\,\textrm{nm}$, relative to the phase-matched $m_{\textrm{ideal}}$ required for optimal fiber-taper coupling. The mode volume $V_{\textrm{eff}}$ is calculated by assuming the fundamental modes to be standing waves, and the TE$_{40}$ to be a traveling wave.}
	\begin{tabular}{c | c | c | c}
		\hline
		Mode & $\lambda_{0}\,\big(\text{nm}\big)$ & $m_{\textrm{pm}}$\,/\,$m$ & $V_{\textrm{eff}}\,\big(\frac{\lambda_0}{n}\big)^{3}$\\ 
		\hline \hline
		TM$_{00}$ & 738 & 0.58 & 29 \\
		TE$_{00}$ & 742 & 0.59 & 33 \\
		TE$_{40}$ & 741 & 0.91 & 84 \\
		\hline
	\end{tabular}
	\label{tbl:phiMatch}
\end{table}

\subsection{Fiber Coupling}
The preferential fiber-taper coupling to the higher-order modes compared to the fundamental modes can be explained  by their relative differences in phase matching to the fiber-taper. For fiber-taper position defined by radial, $\rho$, and azimuthal, $\phi$, coordinates with respect to the microdisk center, a taper with propagation wavenumber $\beta_f$ along the taper axis has electric field varying as  $\exp(-i\beta_f\rho \sin\phi$)\,\cite{Borselli2006}. For a microdisk with a mode field varying as $\exp(im\phi)$, the phase-matching requirement is $m_{\textrm{pm}} \sim \beta_f \rho$\,\cite{Borselli2006}.

Assuming the fiber-taper has the waist radius expected for a single-mode fiber at $640\,\textrm{nm}$\,\cite{Yariv2006}, and evaluating the approximate $\beta_f$ for the wavelengths studied here at $\sim740\,\textrm{nm}$ an $m_{\textrm{pm}}$ of $\sim$27 is required for ideal phase-matching\,\cite{Yariv2006}. Table\,\ref{tbl:phiMatch} demonstrates that, while the fundamental modes show nearly twice the desired angular momentum, the higher-order mode identified through FSR matching has $m$ close to $m_{\textrm{pm}}$ and therefore, favourable phase matching could explain the preferential coupling. While it is not uncommon to couple to microdisk modes demonstrating $m\sim2\times m_{\textrm{pm}}$\,\cite{Borselli2006}, in such instances the devices are often made extremely thin to relax the phase-matching requirements by boosting the intensity of the microdisk field that interacts with the fiber-taper. Table\,\ref{tbl:RelThick} shows that the normalized microdisk thickness, $\tilde{t}=t/\lambda$, of the microdisks studied here are over twice the value of other demonstrated microdisk devices. This is consistent with the stringent phase-matching requirements limiting the coupling to lower-order modes in our system. Finally, Table\,\ref{tbl:phiMatch} shows the difference in $V_{\textrm{eff}}$, defined by the peak field intensity $|E_{\textrm{max}}|$, expected for the fundamental and TE$_{40}$ modes. Here we assume that the fundamental modes are standing waves while the TE$_{40}$ mode is a traveling wave, as observed in measurements of $R_1$ (likely fundamental) and $R_2$ (TE$_{40}$), which are found to be a doublet and a singlet resonance, respectively. The variations in $V$ for mode $j$, and the consequential change in $s_j$, combined with the aforementioned differences in phase matching between the different microdisk modes and the fiber-taper is reflected by the observed discrepancy in $\alpha_{\textrm{I}}$ (see Table\,\ref{tbl:alpha}).

\begin{table}
	\centering
	\caption{Comparison of device thickness normalized to the wavelength of study for several experimentally demonstrated microdisk geometries.}
	
	\begin{tabular}{c | c | c |c} 
	\hline
		Material & $\lambda_0$\,(nm) &$t$\,(nm) & $\tilde{t}$ $\left(\frac{\lambda_0}{n}\right)$ \\
		\hline \hline
		Silicon\,\cite{Borselli2004} & 1550 & 344 & 0.8 \\
		GaP\,\cite{Barclay2009} & 700 & 250 & 1.2\\ 
		Diamond\,\cite{Mitchell2016} & 1550 & 940 & 1.4\\ 
		Diamond (this paper) & 735 & 850 & 2.8\\ 
	\hline
	\end{tabular}
	\label{tbl:RelThick}
\end{table}

\section{Outlook: Optimized microdisk geometry for enhanced emitter-photon coupling}
\begin{figure}[t!]
	\centering\includegraphics[width=\textwidth]{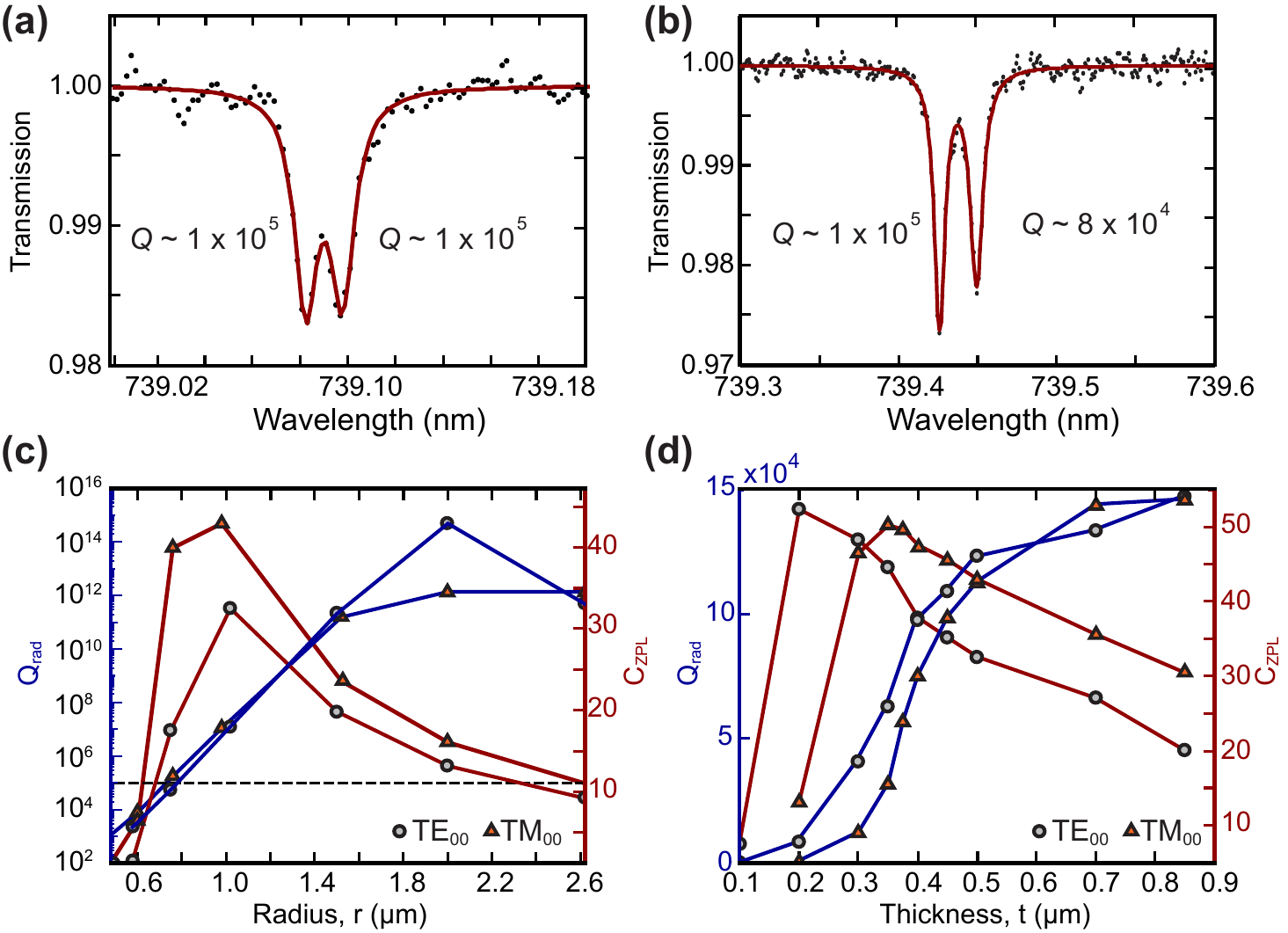}
	\caption{Laser transmission spectroscopy of a diamond microdisk with \textbf{(a)} [$r,t$]=[3.5,\,1.0]\,$\upmu\textrm{m}$ and \textbf{(b)} [$r,t$]=[3.6,\,1.0]\,$\upmu\textrm{m}$. A Lorentzian fit to the cavity resonances reveal $Q\sim1\times10^5$ for both devices. This is the largest observed $Q$-factor for a SCD device operating at visible wavelengths. \textbf{(c)} Simulation of $Q_\text{rad}$ (blue) and $C_{\textrm{ZPL}}$ (burgundy) for the fundamental TE\,(TM)$_{00}$ modes as a function of radius $r$ for a microdisk with fixed $t\sim0.5\,\upmu\textrm{m}$. The dashed horizontal line represents $Q_\text{fab}=1\times10^{5}$, as extracted from the fits in (a) and (b). \textbf{(d)} Simulated $Q_\text{rad}$ (blue) and $C_{\textrm{ZPL}}$ (burgundy) as a function of thickness for a device with fixed $r\sim1\,\upmu\textrm{m}$. The colored lines in (c) and (d) provide a guide to the eye.}
	\label{fig:Q}
\end{figure}

Utilizing the Purcell effect\,\cite{Purcell1946} provides a means to greatly enhance the flux of coherent photons from color centers in diamond by resonant coupling of the ZPL to a single mode in an optical resonator\,\cite{Riedel2017}. For an NV center with lifetime limited optical emission linewidth, optimally positioned in the microdisk with the dipole moment aligned perfectly with the optical field, the Purcell factor $C_{\textrm{ZPL}}$ is given by\,\cite{Janitz2020,Flagan2022}:
\begin{equation}
C_{\textrm{ZPL}} = \zeta_{\textrm{ZPL}}\frac{3}{4\pi^{2}}\frac{Q}{V_{\textrm{eff}}}\left(\frac{\lambda_0}{n}\right)^3\,,
\end{equation}
where $\zeta_{\textrm{ZPL}}=0.03$ is the NV center Debye-Waller factor describing the branching into the ZPL\,\cite{Riedel2017,Santori2010}. The scaling $C_{\textrm{ZPL}}\propto Q/V$ favours the use of fundamental cavity modes with low $m$, on account of their large $Q$-factor and small $V$ (see Table\,\ref{tbl:RelThick})\,\cite{Janitz2020,Shandilya2022}. In Fig.\,\ref{fig:Q}\,(a) and (b), we show laser transmission measurement of diamond microdisks with [$r,t$]=[3.5,\,1.0]\,$\upmu\textrm{m}$ and [$r,t$]=[3.6,\,1.0]\,$\upmu\textrm{m}$, respectively. To extract the $Q$-factor, we fit the cavity resonance with a double Lorentzian, and find $Q\sim1\times10^{5}$ at $\lambda\sim739\,\textrm{nm}$ for both devices. These are, to the best of our knowledge, the largest  $Q$-factors for SCD resonators at visible wavelength reported to date. Here we focus on modes around $\sim737\,\textrm{nm}$ excited by the available laser and resonant with the SiV center\,\cite{Hepp2014}. We expect modes at $637\,\textrm{nm}$, resonant with the NV center ZPL, to have comparable $Q$, depending on the trade-off between mode confinement and surface scattering -- both the sensitivity to surface scattering and the mode confinement increases with shorter wavelengths\,\cite{Borselli2004,Borselli2005}.

We now turn to discuss a potential route to improve the $Q/V$ ratio for our devices. For a general optical resonator, the total $Q$-factor is given by:
\begin{equation}
\frac{1}{Q_{\textrm{tot}}}=\frac{1}{Q_\text{rad}} + \frac{1}{Q_\text{mat}} + \frac{1}{Q_\text{fab}}\,,
\label{eq:Purcell}
\end{equation}
where $Q_{\textrm{rad}}$ accounts for radiation loss, $Q_\text{mat}$ is associated with material absorption and $Q_\text{fab}$ encompasses surface scattering and absorption resulting from fabrication imperfections\,\cite{Borselli2005,Mitchell2019}. Reducing the microdisk dimensions constitutes a method to simultaneously improve the coupling to lower-order fundamental modes and decreasing $V$. Although $Q_\text{rad}$ is expected to decrease with microdisk $r$ and $t$, in this work, we expect $Q_{\textrm{fab}}$ to be the dominant limitation to $Q_{\textrm{tot}}$. As the relative contribution from fabrication induced imperfections in degrading $Q_{\textrm{tot}}$ increases with smaller disk size, $1\times 10^{5}$ (Fig.\,\ref{fig:Q}\,(a) and (b)) provides a realistic upper-limit on $Q_\text{fab}$ expected with the current state-of-the-art fabrication techniques\,\cite{Mitchell2019}. Therefore, in the following analysis, we set $Q_{\textrm{fab}}=10^{5}$ as the upper limit to $Q_{\textrm{tot}}$ i.e.\ $Q_{\textrm{tot}}^{-1}=Q_{\textrm{rad}}^{-1}+10^{-5}$, where we have  ignored $Q_{\textrm{mat}}$ on the grounds of the low absorption in diamond.

In Fig.\,\ref{fig:Q}\,(c) and (d), we simulate the behaviour of $Q_\text{rad}$ as a function of $r$ and $t$, respectively. As expected, we observe an increase in $Q_{\textrm{rad}}$ with increasing $r$ and $t$ (blue lines in Fig.\,\ref{fig:Q}\,(c) and (d)). Furthermore, as indicated by the horizontal dashed line in Fig.\,\ref{fig:Q}\,(c), $Q_{\textrm{rad}}$ only becomes the dominant limitation for $Q_{\textrm{tot}}$ for $r<0.8\,\upmu\textrm{m}$ -- significantly smaller than the range of $r$ for the microdisks studied in this work. This observation supports our assumption that $Q_\text{fab}$ is the dominant limitation to $Q_{\textrm{tot}}$.

Assuming we can maintain the current $Q_\text{fab}$, we next estimate the Purcell factor, $C_{\textrm{ZPL}}$,  expected for our devices at smaller $r$ and $t$. The burgundy lines in Fig.\,\ref{fig:Q}\,(c) and (d) evaluates the theoretical $C_{\textrm{ZPL}}$ according to Eq.\,\ref{eq:Purcell} as a function of the microdisk radius and thickness, respectively. As can be seen, $C_{\textrm{ZPL}}$ would be optimized for a microdisk with $r\sim1\,\upmu\textrm{m}$ and $t\sim400\,\textrm{nm}$, for which a maximum value  $C_{\textrm{ZPL}}\sim50$ is predicted. At this thickness, the devices will have $\tilde{t}=1.4\,\lambda$ -- comparable to what is achieved with the current devices measured at telecom wavelengths\,\cite{Mitchell2016}. 

Select literature values for $C_{\textrm{ZPL}}$, as well as the emitter-independent value $C=C_{\textrm{ZPL}}/\,\zeta_{\textrm{ZPL}}$, are presented in Table\,\ref{tbl:Cliterature}. The results presented in Fig.\,\ref{fig:Q}\,(d) demonstrates that, provided fabrication-related losses can be controlled, diamond microdisks have the ability to match the $C_{\textrm{ZPL}}$ achievable with other state-of-the-art on-chip devices, while providing a broader mode spectrum and direct low-loss coupling without integrated fiber couplers\,\cite{Michael2007,Burek2017}.
\begin{table}
	\centering
	\caption{Expected Purcell factors for SCD devices calculated from the cavity properties as reported in the selected literature.}
	\begin{tabular}{c | c | c | c |c}  
		\hline
		Ref. & Geometry & Material &$C$ &$C_{\textrm{ZPL}}$\\
		\hline \hline
		\cite{Barclay2009} & Microdisk & GaP on diamond & 20 &  0.6\\ 
		\cite{Faraon2012} & Photonic crystal cavity & Diamond & 260 &  7.8\\ 
		\cite{Burek2017} & Photonic crystal cavity & Diamond & 1600 & 48\\
		\cite{Mouradian2017a} & Photonic crystal cavity & Diamond &1300 &39 \\ 
		This Work & Microdisk & Diamond &  1670 & 50
	\end{tabular}
	\label{tbl:Cliterature}
	
\end{table}

\section{Conclusion}
We have performed optical spectroscopy of single-crystal diamond microdisks using fiber-taper-collected PL from an ensemble of NV centers. The relative intensities of resonances in the PL signal compared to those predicted from transmission measurement of mode properties suggest that the PSB emission is filtered not only by the cavity spectrum but also by the mode-dependent coupling between the microdisk and the fiber-taper. A comparison between the measured and simulated FSR implies that, for the devices studied here, there is preferential fiber coupling to higher-order radial modes of relatively high $Q$-factor ($10^{3}-10^{4}$). This preferential coupling is likely a consequence of a more favourable phase-matching conditions. The relatively poor coupling to the fiber-taper limits the characterization of the highest-$Q$ mode families with fiber collection. However, thinning of the microdisks to $\tilde{t}\sim1.4\lambda$ should simultaneously improve the phase-matching-limited fiber-taper coupling and reduce the multi-mode nature of the microdisks by freezing out higher-order vertical modes. If the highest measured $Q_\text{fab}\sim 1\times10^{5}$ can be achieved in such small devices, diamond microdisks should be able to achieve Purcell factors $C_{\textrm{ZPL}}\sim50$, matching those of other state-of-the-art devices while providing broader mode spectrum and alternative integrated coupling options.

\section*{Acknowledgments}
This work was supported by funding from the National Sciences and Engineering Research Council of Canada (NSERC), Alberta Innovates - Technology Futures (AITF) and National Research Council Nanotechnology Research Centre NanoInitiative. SF acknowledges support from the Swiss National Science Foundation (Project No. P500PT\_206919).


\end{document}